\documentclass[twocolumn]{jpsj2}

\title{NEW VACUUM OF BETHE ANSATZ SOLUTIONS IN THIRRING MODEL}

\author{
Takehisa \textsc{Fujita}$^{1}$
\thanks{E-mail address: fffujita@phys.cst.nihon-u.ac.jp}, 
Makoto \textsc{Hiramoto}$^{1}$
\thanks{E-mail address: hiramoto@phys.cst.nihon-u.ac.jp}, 
Takeshi \textsc{Homma}$^{3}$
\thanks{E-mail address: hommaj@phys.cst.nihon-u.ac.jp} and 
Hidenori \textsc{Takahashi}$^{4}$
\thanks{E-mail address: htaka@phys.ge.cst.nihon-u.ac.jp}}

\inst{$^{1}$$^{3}$$^{4}$Department of Physics, Faculty of Science and Technology, 
Nihon University, Tokyo, Japan 
}

\abst{We find a new vacuum of the Bethe ansatz solutions in the massless 
Thirring model. This vacuum breaks the chiral symmetry and 
has the lower energy than the well-known symmetric vacuum energy. 
Further, we evaluate the energy spectrum of the one particle-one hole 
($1p-1h$) states, and find that it has a finite gap. 
The analytical expressions for the true vacuum as well as for the lowest $1p-1h$ 
excited state are also found. Further, we examine the bosonization 
of the massless Thirring model and prove that the well-known procedure 
of bosonization of the massless Thirring model is incomplete 
because of the lack of the zero mode in the boson field. }

\kword{Bethe ansatz, symmetry broken vacuum, zero mode}

\begin{document}
\maketitle

\section{Introduction} 
Symmetries and their breaking have been one of the most 
important subjects in quantum field theory. Since the vacuum can break  
the symmetry of the field theory model, one learns 
the structure of the vacuum and its dynamics of the model through the 
symmetry breaking phenomena \cite{q1,q2,q6}. 

In two dimensions, however, the symmetry breaking in the field theory  
is considered to be different from the four dimensional field theory models. 
In particular, Coleman \cite{q3} presented the proof  
that the two dimensional field theory models cannot spontaneously break the symmetry 
even though the vacuum state may prefer the symmetry broken state. 
However, his proof of the nonexistence of the spontaneous symmetry breaking 
in two dimensions is essentially based on the Goldstone theorem. 
The Goldstone theorem \cite{q1,q2} states that the spontaneous symmetry breaking should 
accompany a massless boson when the vacuum prefers the broken symmetric 
state. However, the massless boson cannot exist in two dimensions since 
it cannot propagate due to the infra-red singularity of the propagator. 
Since this non-existence of the massless boson should hold  rigorously, 
it naturally means that the spontaneous symmetry breaking 
should not occur in two dimensions as long as the Goldstone theorem is right. 

Coleman's theorem looks reasonable, and indeed until recently 
it has been believed to hold true for fermion field theory models as well.  

However, the recent work on the massless Thirring model shows 
that the chiral symmetry of the massless Thirring model is 
spontaneously broken by the Bogoliubov vacuum state \cite{q41,p44,q93,q91}. 
There, the energy of  the new vacuum is lower than that of the free vacuum state, 
and it indeed violates the chiral symmetry. It should be noted that, 
in this analysis, there appears no massless boson, and therefore 
it does not contradict the non-existence of the massless boson in two dimensions. 

This claim, however, does not seem to be accepted yet in general since people 
may believe that the Bogoliubov transformation does not have to be exact, and 
therefore there might be some excuse for the symmetry breaking phenomena 
that occurred in the Thirring model.  

In this paper, we present a new discovery of the symmetry broken vacuum 
of the Bethe ansatz solution in the Thirring model, 
and show that the energy of the new vacuum state 
is indeed  lower than that of the symmetric vacuum state 
even though the symmetric vacuum was considered to be the lowest state 
in the Thirring model. The new vacuum state breaks the chiral symmetry, 
and becomes a massive fermion field theory model. 

Further, we evaluate the energy spectrum of the one particle-one hole states, 
and show that the excitation spectrum has indeed a finite gap. This gap energy 
turns out to be consistent with the effective fermion mass deduced  
from the momentum distribution of the negative energy particles 
in the new vacuum state. This confirms the consistency of the calculation 
of the Bethe ansatz solutions in the Thirring model. 

After carrying out the numerical calculations, we get to know that 
the energies of the vacuum  as well as the lowest one particle-one hole 
state can be expressed analytically. This is quite nice 
since we know clearly which of the vacuum state is the lowest. 
Also, in the thermodynamic limit, the lowest one particle-one hole 
state can be reduced to the effective fermion mass $M_N$ which is described 
in terms of the cutoff $\Lambda$. 

It turns out that there is no massive boson in the Bethe ansatz solutions, 
contrary to the prediction of the Bogoliubov transformation method \cite{q93,q9}. 
However, qualitative properties of the symmetry breaking phenomena 
between the Bethe ansatz calculations and the Bogoliubov method agree with 
each other.

Even though the  Bethe ansatz calculations confirm that there is 
no massless boson in the massless Thirring model, 
some people may claim that the massless 
Thirring model can be bosonized and is reduced to a massless boson 
hamiltonian. Here, we show that the well-known procedure of bosonization 
of the massless Thirring model is incomplete because the zero mode 
of the boson field cannot be defined and quantized. In other words, 
the zero mode of the field $\Phi (0)$ identically vanishes 
in the massless Thirring model. This is in contrast to the Schwinger model 
in which one finds the zero mode of the field $\Phi (0)$ 
by the gauge field $A^1$. Also, it is interesting to note that the massive 
Thirring model has the zero mode through the mass term, and this clearly 
indicates that the massless limit of the massive Thirring model is indeed 
a singular point with respect to the dynamics of the field theory. 

Therefore, the massless Thirring model cannot be reduced to a free 
massless  boson even though it has a similar mathematical structure to 
the massless boson. The spectrum of the massless Thirring model has a finite gap,  
and this is consistent with the fact that there should not be any 
physical massless boson in two dimensions. Even though the defect 
of the bosonization of the massless Thirring model is only one 
point of the boson field, that is, zero mode, it is interesting 
and surprising that $nature$ knows it in advance.  

This paper is organized as follows. In the next section, we discuss 
the Bethe ansatz solutions of the massless Thirring model. We obtain 
the analytic expressions for the vacuum energy as well as for the one 
particle one hole state excitation energy, and show that the new vacuum 
state breaks the chiral symmetry and has the lower energy 
than the symmetric vacuum state. In section 3, we 
present a critical review of the bosonization of the massless 
Thirring model, and show that the massless Thirring model cannot 
be bosonized properly and it has a spectrum with a finite gap. 
In section 4, we summarize what we clarify in this paper. 

\section{Thirring model and Bethe ansatz solutions}
The massless Thirring model is a 1+1 dimensional field theory with 
current current interactions \cite{q7}. 
Its hamiltonian can be written as 
$$H = \int dx \left\{-i\left(\psi_1^{\dagger}{\partial\over{\partial x}}\psi_1
-\psi_2^{\dagger}{\partial\over{\partial x}}\psi_2 \right)
+2g \psi_1^{\dagger}\psi_2^{\dagger}\psi_2\psi_1 \right\}.  \eqno{(2.1)}$$
The hamiltonian eq.(2.1) can be diagonalized by the Bethe ansatz wave 
function for $N$ particles  \cite{q16,q7,q42}
$$\mid k_1, \cdots, k_N \rangle =\int dx_1 \cdots dx_{N_1}dy_1 \cdots dy_{N_2} $$
$$ \times \prod_{i=1}^{N_1}  \exp (ik_ix_i) 
\prod_{j=1}^{N_2}  \exp \left(ik_{N_1+j}y_j\right) $$
$$ \times \prod_{i,j}\left(1+\lambda \theta (x_i-y_j)
 \right) \prod_{i=1}^{N_1}\psi_1^{\dagger}(x_i)
\prod_{j=1}^{N_2}\psi_2^{\dagger}(y_j)
\mid \! 0 \rangle, \eqno{(2.2)} $$
with $N_1+N_2 =N$. $\theta (x)$ denotes the step function. 
$k_i$ represents the momentum of the $i-$th particle. 
$ \lambda $ is determined to be \cite{q16}
$$ \lambda = -{g\over 2} S_{ij}  \eqno{(2.3)} $$
where $S_{ij}$ is defined as
$$ S_{ij}={k_iE_j-k_jE_i\over {k_ik_j-E_iE_j-\epsilon^2 }} . \eqno{(2.4)} $$ 
Here, $\epsilon$ denotes the infra-red regulator which should be infinitesmally small.   
For the infra-red regulator $\epsilon$, it is important to note 
that the physical observables like momnetum $k_i$ do not depend on the regulator $\epsilon$. 
The derivation of eq.(2.3) is given in Appendix. 

In this case, the eigenvalue equation becomes 
$$ H \mid k_1, \cdots, k_N \rangle  = \sum_{i=1}^N E_i  \mid
 k_1, \cdots, k_N \rangle  . \eqno{(2.5)} $$ 
From the periodic boundary condition (PBC), one obtains the following PBC equations,
$$ k_i = {2\pi n_i \over L} + {2 \over L}\sum_{j\not= i}^N 
\tan^{-1} \left({g\over 2} S_{ij} \right) \eqno{(2.6)} $$
where $n_i$'s are integer, and runs as $n_i=0, \pm 1,\pm 2, \cdots , N_0$ 
where
$$ N_0={1\over 2}(N-1). $$  

\subsection{Vacuum state}

First, we want to make a vacuum.  We write the PBC equations
for the vacuum which is filled with negative energy particles \cite{q13,q133}
$$ k_i = {2\pi n_i \over L} - {2 \over L}\sum_{\stackrel{\scriptstyle i\ne j}{k_i\ne k_j}}^N 
 \tan^{-1} \left({g\over 2} {k_i|k_j|-k_j|k_i|\over{k_ik_j-|k_i||k_j|-\epsilon^2 }} 
\right). \eqno{(2.7)} $$
Although, the expression of $S_{ij}$ is different from that of
Odaka and Tokitake\cite{q42}, Andrei and Lowenstein \cite{q43}, 
it produces the same values of the Bethe ansatz solutions of the symmetric vacuum state. 
Here, we first fix the maximum 
momentum of the negative energy particles, 
and denote it by the cut off momentum $\Lambda$. Next, 
we take the specific value of $N$, and this leads to the determination of $L$ 
$$ L={2\pi N_0\over{\Lambda}} . \eqno{(2.8)} $$
If we solve eq.(2.7), then we can determine the vacuum state, 
and the vacuum energy $E_v$  can be written as
$$  E_v =- \sum_{i=1}^{N} |k_i| . \eqno{(2.9)} $$
It should be noted that physical observables are obtained by taking 
the thermodynamic limit where we let $L \rightarrow \infty$ and $N \rightarrow \infty$, 
keeping $\Lambda$ finite. If there is other scale like the mass, then one should 
take the  $\Lambda$ which is sufficiently large compared to the other scale. 
However, there is no other scale in the massless Thirring model or four dimensional QCD 
with massless fermions, and therefore all the physical observables are measured 
by the  $\Lambda$. Here, we can take all the necessary steps, if required,  since 
all the physical quantities are given analytically. In fact, as we see below, 
the excitation energy and the effective fermion mass are expressed in terms of 
the  $\Lambda$ in the thermodynamic limit. 

\subsection{Symmetric vacuum state}

The solution of eq.(2.7) has been known and is written as \cite{q7,q42} 
$$ k_1=0 \eqno{(2.10a)} $$
for $n_1=0$,      
$$ k_i = {2\pi n_i \over L} + {2N_{0} \over L}
\tan^{-1} \left({g\over 2}  \right) \eqno{(2.10b)} $$
for $n_i=1,2,\cdots, N_0$,     
$$ k_i = {2\pi n_i \over L} - {2N_{0} \over L}
\tan^{-1} \left({g\over 2}  \right) \eqno{(2.10c)} $$
for $n_i=-1,-2,\cdots, -N_0$. This gives a symmetric vacuum state, 
and was considered to be the lowest state. 

The vacuum energy $E_v^{\rm sym}$ can be written as 
$$ E_v^{\rm sym}=-\Lambda \left\{ N_0+1+{2N_0\over \pi}
\tan^{-1} \left({g\over 2} \right) \right\} . \eqno{(2.11)} $$

\subsection{True vacuum state}

It is surprising that eq.(2.7) has a completely different solution from the above 
analytical solutions. By the numerical calculation of eq.(2.7), 
we first find the new vacuum state. After that, we get to know that the solutions 
can be analytically written like the symmetric case,
$$ k_1={2N_0 \over L} \tan^{-1} \left({g\over 2}  \right) \eqno{(2.12a)} $$
for $n_1=0$,      
$$ k_i = {2\pi n_i \over L} + {2N_{0} \over L}
\tan^{-1} \left({g\over 2}  \right) \eqno{(2.12b)} $$
for $n_i=1,2,\cdots, N_0$,     
$$ k_i = {2\pi n_i \over L} - {2(N_{0}+1) \over L}
\tan^{-1} \left({g\over 2}  \right) \eqno{(2.12c)} $$
for $n_i=-1,-2,\cdots, -N_0$. The new vacuum  has no $k_i=0$ solution, 
and breaks the left-right symmetry.  Instead, all of the momenta of the 
negative energy particles become finite. 

The energy $E_v^{\rm true}$ of the true vacuum state can be written as 
$$ E_v^{\rm true}=-\Lambda \left\{ N_0+1+{2(N_0+1)\over \pi}
\tan^{-1} \left({g\over 2} \right) \right\} . \eqno{(2.13)} $$

From the distributions of the negative energy 
particles, one sees that this solution breaks the chiral symmetry. 
This situation can be easily seen from the analytical solutions 
since the absolute value of the momentum of the negative energy particles is 
higher than $\displaystyle{{\Lambda \over \pi} \tan^{-1} \left({g\over 2}  \right)} $. 
Therefore, we can define the effective fermion mass $M_N$ by 
$$ M_N = {\Lambda \over \pi} \tan^{-1} \left({g\over 2}  \right) . 
\eqno{(2.14)}  $$
In Table \ref{t1}, we show the calculated results of the new vacuum as well as 
the symmetric vacuum energies as the function of  the particle 
number $N$. Here, we present the case with the coupling constant of $g= \pi $.  

\begin{table}[tb]
\caption{We show the calculated results of the vacuum energy of Bethe ansatz solutions 
at $ g=\pi $ with the particle number $N=401$ and $N=1601$.  
${\cal{E}}^{\rm sym}_v$  and  ${\cal{E}}^{\rm true}_v$ denote the symmetric vacuum 
and the true vacuum energies, respectively. 
We also show the effective fermion mass ${\cal M}_N$ deduced from the vacuum momentum 
distributions. All the energies are measured in units of $\Lambda$, 
namely, ${\cal E} \equiv E/\Lambda $ and ${\cal M}_N \equiv M_N/\Lambda $.}
\label{t1}

\vspace{4mm}
\begin{tabular}{cccc}
\hline
\ $N$ & ${\cal{E}}^{\rm sym}_v $  & ${\cal{E}}^{\rm true}_v$  &  ${\cal M}_N$ \\
\hline
401 & $-328.819$ & $-329.458$ & $0.320$  \\
\hline
1601 & $-1312.274$    & $-1312.913$ & 0.320  \\
\hline
\end{tabular} 
\end{table}

\subsection{$1p-1h$ state}

Next, we evaluate one particle-one hole $(1p-1h)$
 states. There, we take out one
negative energy particle ($i_0$-th particle)
 and put it into a positive energy state.
In this case, the PBC equations become
$$  k_i  = \frac{2\pi n_i}{L}-\frac{2}{L}\tan^{-1}
 \left( {g\over 2} {k_i|k_{i_0}|+k_{i_0}|k_i|\over{k_ik_{i_0}+|k_i||k_{i_0}| +\epsilon^2 }}  \right) $$
$$
 - \frac{2}{L}\sum_{\stackrel{\scriptstyle j\neq i,i_0}{k_j\ne k_i,k_{i_0}}}^N\tan^{-1}
\left({g\over 2} {k_i|k_j|-k_j|k_i|\over{k_ik_j-|k_i||k_j| -\epsilon^2 }} 
\right) \eqno{(2.15a)}$$
for $i\neq i_0$. $$  k_{i_0}  = \frac{2\pi n_{i_0}}{L}-
\frac{2}{L}\sum_{\stackrel{\scriptstyle j\neq i_0}{k_j\ne -k_{i_0}}}^N
 \tan^{-1}\left({g\over 2} {k_{i_0}|k_j|+k_j|k_{i_0}|
\over{k_{i_0}k_j+|k_{i_0}||k_j| +\epsilon^2 }} \right) \eqno{(2.15b)}$$
for $i= i_0$. In this case, the energy of the one particle-one
hole states $E^{1p1h}_{(i_0)}$ is given as,
$$  E^{1p1h}_{(i_0)}= |k_{i_0}| -
\sum_{\stackrel{\scriptstyle i=1}{i\not= i_0}}^{N}
|k_i|  .  \eqno{(2.16)}   $$
It turns out that the solutions of eqs.(2.15) can be found 
at the specific value of $n_{i_0}$ and then from this $n_{i_0}$ value  on, 
we find continuous spectrum of the $1p-1h$ states. 

Here, we show the analytical solution of eqs.(2.15) for the lowest $1p-1h$ state. 
$$ k_{i_0}={2\pi n_{i_0}\over L}-{2N_0\over L}\tan^{-1}\left({g\over 2}\right) \eqno{(2.17a)} $$  
for $n_{i_0}$,      
$$ k_i = {2\pi n_i \over L} + {2(N_{0}+1) \over L}
\tan^{-1} \left({g\over 2}  \right)\eqno{(2.17b)} $$
for $n_i=0,1,2,\cdots, N_0$    
$$ k_i = {2\pi n_i \over L} - {2N_{0} \over L}
\tan^{-1} \left({g\over 2}  \right)\eqno{(2.17c)} $$
for $n_i=-1,-2,\cdots,-N_0$. $n_{i_0}$ is given by 
$$ n_{i_0}=\left[ {N_0\over{\pi}}\tan^{-1} \left({g\over 2}\right)  \right] , 
\eqno{(2.18)} $$ 
where $[X]$ denotes the smallest integer value which is larger than $X$. 
In this case, we can express the lowest $1p-1h$ state energy analytically
$$ E_0^{1p-1h}=-\Lambda
\left\{(N_0+1)-{{2n_{i_0}}\over N_0}+{2(N_0+1)\over \pi}\tan^{-1}\left({g\over\pi}\right)
\right\}. \eqno{(2.19)} $$
Therefore, the lowest excitation energy $\Delta E_0^{1p-1h}$ 
with respect to the true vacuum state becomes 
$$ \Delta  E_0^{1p-1h} \equiv E_0^{1p-1h}-E_v^{\rm true}=
{2\Lambda\over N_0}  n_{i_0} . \eqno{(2.20)} $$
If we take the thermodynamic limit, that is, $ N\rightarrow \infty $ and  
$ L\rightarrow \infty $, then eq.(2.18) can be reduced to 
$$ \Delta  E_0^{1p-1h} ={2\Lambda \over \pi} \tan^{-1} 
\left({g\over 2}  \right)  = 2M_N  . \eqno{(2.21)} $$
In Table \ref{t2}, we show the lowest five states of the $1p-1h$ energy by the numerical 
calculation. From this, we can determine the gap energy. 

\begin{table}[tb]
\caption{We show several lowest states of the calculated results of the 1p-1h states energy 
$\cal{E}$ of eqs.(2.19) at $ g=\pi  $ with $N=1601$. 
The gap energy $\Delta {\cal E} \equiv {\cal E}^{(1p1h)}-{\cal E}_v$ is also shown. 
All the energies are measured in units of $\Lambda$.}
\label{t2}

\vspace{4mm}
\begin{tabular}{ccc}
\hline
\   & $\cal{E}$  &  $\Delta \cal{E} $ \\
\hline
 vacuum  & $-1312.913$  & \   \\
\hline
${1p-1h}$ (0) & $-1312.273$ & $0.640$  \\
\hline
${1p-1h}$ (1)  & $-1312.272$ & $0.641$  \\
\hline
${1p-1h}$ (2)  & $-1312.271$ & $0.642$  \\
\hline
${1p-1h}$ (3)   & $-1312.269$ & $0.644$  \\
\hline
${1p-1h}$ (4)  & $-1312.268$ & $0.645$  \\
\hline
\end{tabular} 
\end{table}

From this gap energy, we can obtain the effective fermion mass which is one half 
of the lowest gap energy. This can be easily given as 
$$ M_N = 0.320 \  \Lambda . \eqno{(2.22)}  $$
This is consistent with the effective fermion mass 
deduced from the negative energy distribution of the vacuum. 
This confirms the consistency of the present calculations. 

\subsection{Boson state}

In this calculation, we do not find any boson state, contrary to the prediction 
of the Bogoliubov transformation method. Since the present calculation is exact, 
we believe that the Bogoliubov calculation overestimates the attraction 
between the particle hole states. The main difference between the Bethe solutions and 
the Bogoliubov vacuum arises from the dispersion relation 
of the negative energy particles. From the Bethe ansatz solutions, 
it is clear that one cannot make a simple free particle dispersion 
with the fermion mass term while the Bogoliubov method 
assumes the free fermion dispersion relation for the negative energy particles. 
This should generate slightly stronger attraction for the Bogoliubov vacuum state 
than for the Bethe ansatz solution. 

However, as far as the symmetry breaking mechanism is concerned, the Bogoliubov 
transformation gives a sufficiently reliable description of the dynamics 
in the spontaneous symmetry breaking phenomena. 

\section{Bosonizations}

Here, we briefly review the bosonization procedure in two dimensional 
field theory models. In particular, we discuss the Schwinger model and 
the massless and massive Thirring models and show that the massless 
Thirring model cannot be bosonized properly due to the lack 
of the zero mode of the boson field. 

\subsection{Schwinger model}

The best known model of the bosonization is the Schwinger model \cite{p1} 
which is the two dimensional QED with massless fermions. 
In the Schwinger model, one takes a Coulomb gauge, and 
in this case, the space part of the  vector potential $A^1$ depends on 
time and corresponds to the zero mode of the boson field \cite{p2}. 
If one defines the fermion current $j_{\mu}=\bar{\psi}\gamma_{\mu}\psi $, 
then the momentum representation $\tilde{j_{\mu} }$ of the current  is 
related to the boson field and its conjugate field as 
$$ \tilde{j_{0} }(p) =ip\sqrt{L\over \pi} \Phi (p) \qquad {\rm for}\ \ \  p 
\not= 0 \eqno{(3.1a)} $$
$$ \tilde{j_{1} }(p) =\sqrt{L\over \pi} \Pi (p) \qquad {\rm for}\ \ \  p \not= 0  \eqno{(3.1b)} $$
where $\Phi (p)$ and $\Pi (p)$ denote the boson field and its 
conjugate field, respectively. $L$ denotes the box length. 

It is very important to note that $\Pi (0)$ and $\Phi (0)$ are not defined in 
eqs.(3.1). In the Schwinger model, they are related to the chiral 
charge and its time derivative as 
$$ \Pi (0) = {\pi\over{g\sqrt{ L}}} Q_5 \eqno{(3.2a)} $$
$$ \Phi (0) = {\pi\over{g\sqrt{ L}}} {\dot Q}_5 \eqno{(3.2b)} $$
where ${\dot Q}_5 $ is described by the vector field $A^1$ due to 
the anomaly equation 
$$ {\dot Q}_5 ={gL\over \pi} {\dot A}^1 . \eqno{(3.3)} $$
From these identification, one can write down the hamiltonian 
for the Schwinger model
$$ H= \sum_{p} \left\{ {1\over 2} {\Pi}^{\dagger} (p) \Pi (p)
+{1\over 2}p^2 {\Phi}^\dagger (p)\Phi (p) + 
{g^2\over{2 \pi}} {\Phi}^{\dagger} (p) \Phi (p) \right\}. 
\eqno{(3.4)} $$
This is just the free massive boson hamiltonian. 

\subsection{Massless Thirring model}

It has been believed that the massless 
Thirring model can be bosonized \cite{p33,p3} in the same way as above, and its hamiltonian is written 
$$ H= {1\over 2}\sum_{p \not= 0} \left\{ (1-{g\over{2\pi}})  {\Pi}^{\dagger} (p) \Pi (p)
+(1+{g\over{2\pi}}) p^2 {\Phi}^\dagger (p)\Phi (p)  \right\} . 
\eqno{(3.5)} $$
This looks plausible, but one knows at the same time 
that the $p=0$ part is not included. In fact, 
there is a serious problem in the definition 
of the boson field $\Phi (0)$ and $\Pi (0)$ at the zero momentum $p=0$. 
From eqs.(3.1), it is clear that one cannot define the zero mode 
of the boson field. In the Schwinger model, 
one finds the $\Phi (0)$ due to the anomaly equation. However, the Thirring 
model has no anomaly, and therefore the $\Phi (0)$ identically vanishes. 
That is, 
$$ \Phi (0) =0 .  \eqno{(3.6)} $$
There is no way to find the corresponding zero mode of the boson field 
in the massless Thirring model since the axial vector current is always conserved. 

Therefore, the hamiltonian of the massless 
Thirring model eq.(3.5) does not correspond to the massless boson. 
It is interesting to notice 
that the problem is closely related to the zero mode which exhibits 
the infra-red property of the hamiltonian. This is just consistent 
with the non-existence of the massless boson due to the infra-red 
singularity of the propagator in  two dimensions \cite{p4}. 
Further, as discussed in the previous section, the Bethe ansatz solutions 
confirm the finite gap of the massless Thirring spectrum, and 
this rules out a possibility of any excuse of the massless boson 
in the massless Thirring model. 

\subsection{Massive Thirring model}

It is well known that the massive Thirring model is equivalent 
to the sine-Gordon field theory \cite{p7}. 
The proof of the equivalence is 
based on the observation that the arbitrary number of the correlation 
functions between the two models agree with each other 
if some constants and the fields of the two models are properly identified between them.  
This indicates that the massive 
Thirring model must be well bosonized. 

This is now quite clear since the axial vector current conservation 
is violated by the mass term,
$$ \partial_\mu j^{\mu}_5 = 2im \bar \psi \gamma_5  \psi 
\eqno{(3.7)} $$
where $j_{\mu}^5$ is defined as 
$$  j_{\mu}^5 = \bar \psi \gamma_5 \gamma_{\mu} \psi . \eqno{ (3.8)}   $$
It should be noted that the $j_{0}^5$ is equal to $j_{1}$ in two dimensions. 

Therefore, one can always define the ${\dot Q}_5$ by
$$ {\dot Q}_5 =2im \int \bar \psi \gamma_5  \psi dx . \eqno{(3.9)} $$
Therefore, one obtains the field $\Phi (0)$ of the boson 
in terms of eqs.(3.2) and (3.9). 
$$ \Phi (0) = {2im\pi\over{g\sqrt{ L}}} \int \bar \psi \gamma_5  \psi dx . \eqno{(3.10)} $$

\subsection{Physics of zero mode}

What is the physics behind the hamiltonian without the zero mode ? 
Here, we discuss the effect of the zero mode and the eigenvalues of the hamiltonian 
in a simplified way. The hamiltonian eq.(3.5) can be rewritten as 
$$ H= H_B
-{1\over 2} \left( 1-{g\over{2\pi}} \right)  {\Pi}^{\dagger} (0) \Pi (0) \eqno{(3.11)} $$
where the $\Pi(0)$ field is introduced by hand, and the existence of 
the  $\Pi(0)$ and $\Phi(0)$  fields is assumed. 
Here, $H_B$ denotes the free boson hamiltonian and is written as 
$$ H_B= {1\over 2}\sum_{p } \left\{ (1-{g\over{2\pi}})  {\Pi}^{\dagger} (p) \Pi (p)
+(1+{g\over{2\pi}}) p^2 {\Phi}^\dagger (p)\Phi (p)  \right\} . \eqno{(3.12)} $$
Now, we assume the following eigenstates for $H_B$ and 
$ {\Pi}^{\dagger} (0) \Pi (0)$ by
$$ H_B|p\rangle =E_p|p\rangle \eqno{(3.13a)} $$
$$ {\Pi}^{\dagger} (0) \Pi (0) |\Lambda \rangle  =\Lambda |\Lambda \rangle \eqno{(3.13b)} $$
where $E_p={2\pi\over L} p  \ \ {\rm with} \ \ p=0,1,2, \cdots  $, and  
$\Lambda$ is related to the box length $L$ by $\Lambda ={c_0\over{L}}$ with $c_0$ constant. 

Eq. (3.13a) is just the normal eigenvalue equation for the massless boson and 
its spectrum. On the other hand, eq.(3.13b) is somewhat artificial 
since the state $ |\Lambda \rangle $ is introduced by hand. The zero mode state 
of the hamiltonian $H_B$ should couple with the state $ |\Lambda \rangle $, and 
therefore new states can be made by the superposition of the two states
$$  |v \rangle = c_1  |\Lambda \rangle +c_2 
|0 \rangle  \eqno{(3.14)} $$
where $c_1$ and $c_2$ are constants. 
Further, we assume for simplicity that the overlapping integral between the $|0\rangle$ 
and the $|\Lambda \rangle$ states is small and is given by $\epsilon$
 $$ \langle 0|\Lambda \rangle =\epsilon . \eqno{(3.15)} $$
In this case, the energy eigenvalues $\langle v  |H |v \rangle$ of eq.(3.11) 
become at the order of $O(\epsilon)$
$$ E_{\Lambda} = \langle \Lambda  |H_B |\Lambda \rangle 
-{1\over 2} \left( 1-{g\over{2\pi}} \right) \Lambda \eqno{(3.16a)} $$
$$ E_0 = -{1\over 2} \left( 1-{g\over{2\pi}} \right) 
\langle 0  | {\Pi}^{\dagger} (0) \Pi (0)|0 \rangle . \eqno{(3.16b)} $$
If we assume that the magnitude of the $\langle \Lambda  |H_B |\Lambda \rangle $ and 
$\langle 0  | {\Pi}^{\dagger} (0) \Pi (0)|0 \rangle $ should be appreciably smaller than the $\Lambda$, 
$$ \langle \Lambda  |H_B |\Lambda \rangle \ll  \Lambda \eqno{(3.17a)}  $$ 
$$ \langle 0  | {\Pi}^{\dagger} (0) \Pi (0)|0 \rangle  \ll  \Lambda \eqno{(3.17b)}  $$ 
then the spectrum of the hamiltonian eq.(3.11) has a finite gap, and 
the continuum states of the massless excitations start right above the gap. This is 
just the same as the spectrum obtained from the Bethe ansatz solutions discussed 
in the previous section. 

\section{Conclusions}
It has been believed for a long time that the Bethe ansatz solution of the massless 
Thirring model has only a symmetric solution for the vacuum, 
and this symmetric vacuum state has been considered to be the real vacuum since 
it was thought to be the lowest energy state. 

Here, we have presented a symmetry broken vacuum of the Bethe ansatz solutions 
in the Thirring model, and have shown that the true vacuum energy is 
indeed lower than the symmetric vacuum energy. 
This is quite surprising since the symmetry preserving state often 
gives the lowest energy state in quantum mechanics. However, in the field theory model, 
there is also the case in which the symmetry is spontaneously broken in the vacuum 
state \cite{p97}, and this is indeed what is realized and observed in the Thirring model. 

In this new vacuum state, the chiral symmetry is broken, and therefore 
the momentum distribution of the negative energy state become a massive fermion 
theory. From the distribution of the vacuum momentum, we deduce the fermion mass. 

We have also calculated the one particle-one hole excitation spectrum, and 
found that the spectrum has a finite gap. From this gap energy, we can determine 
the fermion mass, and confirm that the fermion mass from the gap 
energy agrees with the one which is estimated from the vacuum momentum distribution.  

Also, we have shown that the bosonization procedure of the massless Thirring 
model has a serious defect since there is no corresponding zero mode 
of the boson field and that the massless Thirring model therefore 
cannot be fully bosonized.  

Since the massless Thirring model cannot be bosonized properly, 
there is no massless excitation spectrum in the model, and 
this is consistent with the Bethe ansatz solutions that the massless 
Thirring model has a finite gap and then the continuum spectrum starts 
right above the gap. 

Also, we should stress that the bosonization of the massless Thirring model 
has a subtlety, and one must be very careful for treating it. If one makes 
a small approximation or a subtle mistake in calculating the spectrum 
of the hamiltonian, then one would easily obtain unphysical 
massless excitations from the massless Thirring model. 
We believe that the same care must be taken for the $SU(N)$ Thirring 
model where some approximations like the $1/N$  expansion are made and 
the massless boson is predicted \cite{p8}. When we discuss the 
large $N$ expansion, there are serious problems related to the $1/N$ 
approximation. The basic point is that they cannot take into 
account the subtlety of the dynamics. In particular, if one makes first 
the large $N$ limit, then one loses some important interactions which contribute 
to the boson mass. As Gross and Neveu pointed out in their paper \cite{q99}, 
the massless boson does not exist if they were to calculate to the higher 
orders in $1/N$. The existence of massless boson will give rise to infrared 
infinities arising from virtual states. This means that the lowest order 
approximation in $1/N$ is meaningless, and to investigate 
the infrared stability of the theory one has to work to all orders in $1/N$. 
This infra-red problems become particularly important when treating the bound state 
like boson mass.  

It is clear by now that the present results are 
in contradiction with Coleman's theorem \cite{q3}. 
In this paper, we have presented counter examples against Coleman's theorem, and 
the exact solutions in two dimensional field theory should correspond 
to "experimental facts". Therefore, one should figure out 
the mathematical reason why Coleman's theorem is violated in fermion 
field theory model. 
In addition to the massless Thirring model, QCD with massless fermions 
in two dimensions spontaneously breaks the chiral symmetry 
with the axial vector current conservation. 
Therefore, the massless QCD$_2$ is also in contradiction 
with Coleman's theorem, but there is no massless boson \cite{q34}, and 
in this sense, it does not violate the theorem that there should not 
exist any massless boson in two dimensions. In reality, there is no example 
of fermion field theory models in which the symmetry of the vacuum state 
is $not$ broken due to Coleman's theorem. 
However, the basic and mathematical problem with Coleman's theorem 
is still unsolved in this paper and is left for the reader. But we believe that 
the basic problem of the symmetry breaking business in two dimensions must 
come from the Goldstone theorem itself for the fermion field theory \cite{q93}.  

At this point, we should make a comment on the correspondence between 
the Thirring model and the Heisenberg $XXZ$ model. It is believed that the two 
models are equivalent to each other. However, it is now clear that the spectrum 
of the Thirring model gives a finite gap while the Heisenberg $XXZ$ model 
predicts always gapless spectrum. This means that, even though the two models 
are mathematically shown to be equivalent to each other, 
they are physically very different \cite{q92}. 

What should be the main reason for the difference ? 
If one makes the field theory into the lattice, then 
the lattice field theory loses some important continuous symmetry like Lorentz invariance 
or chiral symmetry. If the lost symmetry plays some important 
role for the spectrum of the model, then the lattice field theory becomes 
completely a different model from the continuous field theory model \cite{q98}. 
In general, the way of cutting the continuous space into a discrete one is not unique, 
and equal cutting of the space may not be sufficient for some 
of the field theory models. 

The present work is concerned with a specific model in two dimensions, and quite 
different from four dimensional field theory models. However, 
the present work certainly raises a warning on the lattice version 
of the field theory since clearly there are important continuous symmetries 
in any of the field theory models in four dimensions, and if these symmetries 
may be lost in the lattice version, then it is quite probable that the lattice 
calculations may not be able to reproduce a physically important spectrum 
of the continuous field theory models.

\vspace{2cm}

\appendix

\section{Derivation of phase shift function at massless limit}

We present the derivation of eq.(2.4) from the Bethe ansatz equations  
of the massive Thirring model which are given by  Bergknoff and 
Thacker \cite{q16}. 
According to the Bethe ansatz, the Hamiltonian can be digonalized  
when the phase shift function $S_{ij}$ is written as
\begin{equation}
S_{ij}  =  \frac{\sin(\theta_{k_i} - \theta_{k_j})}
{\sin(\theta_{k_i} + \theta_{k_j})}
\end{equation}
where 
\begin{equation}
 \tan 2 \theta_{k_i} = \frac{m_0}{k_i}
\end{equation}
Therefore, we can rewrite as
\begin{subequations}
\begin{align}
\sin \theta_{k_i} &=   \sqrt{\dfrac{1-\cos 2 \theta_{k_i}}{2}} = 
  \dfrac{\sqrt{E_i-k_i}}{\sqrt{2E_i}}, \\
\cos \theta_{k_i} &=   \sqrt{\dfrac{1+\cos 2 \theta_{k_i}}{2}} = 
  \dfrac{\sqrt{E_i+k_i}}{\sqrt{2E_i}},
\end{align}
\end{subequations}
where $E_i= \sqrt{k_i^2+m_0^2}$. 
In this case, $S_{ij}$  becomes 
\begin{align}
S_{ij}  &=  \dfrac{\sqrt{E_i-k_i} \sqrt{E_j+k_j} - \sqrt{E_i+k_i} 
\sqrt{E_j-k_j}}
                     {\sqrt{E_i-k_i} \sqrt{E_j+k_j} + \sqrt{E_i+k_i} \sqrt{E_j-k_j}} 
\notag \\
            & =  \dfrac{k_i E_j - k_j E_i}{k_i k_j - E_i E_j - m_0^2}.
\end{align}
For the massless limit $m_0 \rightarrow 0$, $E_i = \sqrt{k_i^2+m_0^2} \rightarrow |k_i|$. 
Therefore, we have the phase shift function $S_{ij}$  of the \textit{massless} 
Thirring model with the 
\textit{regulator} $\epsilon $ as
\begin{equation}
S_{ij} =  \dfrac{k_i |k_j| - k_j |k_i|}{k_i k_j - |k_i||k_j| 
- \epsilon^2}.
\end{equation}
Here, it should be important to note that the solutions of eq.(2.7) do {\it not}
depend on the regulator $\epsilon$. Therefore, we can take 
the massless limit properly.

\end{document}